\documentclass[]{aastex631}
\usepackage{amsmath}
\usepackage{comment}
\usepackage{textgreek}
\usepackage{multirow}

\shorttitle{}
\shortauthors{Kato et al., 2022}
\graphicspath{{./}{figures/}}

\begin{document}

\title{On the source contribution to the Galactic diffuse gamma rays above 398 TeV detected by the Tibet AS$\gamma$ experiment}

\author{S. Kato}
\altaffiliation{Corresponding author}
\affiliation{Institute for Cosmic Ray Research, University of Tokyo, Kashiwa 277-8582, Japan; katosei@icrr.u-tokyo.ac.jp}
\author{D. Chen}
\affiliation{National Astronomical Observatories, Chinese Academy of Sciences, Beijing 100012, China}
\author{J. Huang}
\affiliation{Key Laboratory of Particle Astrophysics, Institute of High Energy Physics, Chinese Academy of Sciences, Beijing 100049, China} 
\author{T. Kawashima}
\affiliation{Institute for Cosmic Ray Research, University of Tokyo, Kashiwa 277-8582, Japan; katosei@icrr.u-tokyo.ac.jp}
\author{K. Kawata}
\affiliation{Institute for Cosmic Ray Research, University of Tokyo, Kashiwa 277-8582, Japan; katosei@icrr.u-tokyo.ac.jp}
\author{A. Mizuno}
\affiliation{Institute for Cosmic Ray Research, University of Tokyo, Kashiwa 277-8582, Japan; katosei@icrr.u-tokyo.ac.jp}
\author{M. Ohnishi}
\affiliation{Institute for Cosmic Ray Research, University of Tokyo, Kashiwa 277-8582, Japan; katosei@icrr.u-tokyo.ac.jp}
\author{T. Sako}
\affiliation{Institute for Cosmic Ray Research, University of Tokyo, Kashiwa 277-8582, Japan; katosei@icrr.u-tokyo.ac.jp}
\author{T. K. Sako}
\affiliation{Institute for Cosmic Ray Research, University of Tokyo, Kashiwa 277-8582, Japan; katosei@icrr.u-tokyo.ac.jp}
\author{M. Takita}
\affiliation{Institute for Cosmic Ray Research, University of Tokyo, Kashiwa 277-8582, Japan; katosei@icrr.u-tokyo.ac.jp}
\author{Y. Yokoe}
\affiliation{Institute for Cosmic Ray Research, University of Tokyo, Kashiwa 277-8582, Japan; katosei@icrr.u-tokyo.ac.jp}

\begin{abstract}
  Potential contribution from gamma-ray sources to the Galactic diffuse gamma rays observed above 100 TeV (sub-PeV energy range) by the Tibet AS$\gamma$ experiment is an important key to interpreting recent multi-messenger observations. This paper reveals a surprising fact: none of the 23 Tibet AS$\gamma$ diffuse gamma-ray events above $398\, {\rm TeV}$ within the Galactic latitudinal range of $|b|<10^{\circ}$ come from the 43 sub-PeV gamma-ray sources reported in the 1LHAASO catalog, which proves that these sources are not the origins of the Tibet AS$\gamma$ diffuse gamma-ray events. No positional overlap between the Tibet AS$\gamma$ diffuse gamma-ray events and the sub-PeV LHAASO sources currently supports the diffusive nature of the Tibet AS$\gamma$ diffuse gamma-ray events, although their potential origin in the gamma-ray sources yet unresolved in the sub-PeV energy range cannot be ruled out.
    %
\end{abstract}

\keywords{Galactic cosmic rays (567) --- Gamma-ray astronomy (628)}

\section{Introduction} \label{sec:intro}
The Tibet AS$\gamma$ experiment has performed the first detection of Galactic diffuse gamma rays above $100\, {\rm TeV}$ (sub-PeV range) well concentrated on the Galactic Plane \citep{TibetDiffuse}. The hadronic origin of the Tibet AS$\gamma$ diffuse gamma rays is supported by the observation of Galactic neutrinos by IceCube \citep{IceCube_2023}; the energy spectrum of the Tibet AS$\gamma$ diffuse gamma rays converted into the neutrino flux assuming the hadronic interaction smoothly connects at $\simeq 60\, {\rm TeV}$ with the all-sky Galactic-neutrino flux scaled by the intensity of the diffuse emission template proposed by \cite{Ackermann_2012} in the field of view of the Tibet AS$\gamma$ experiment. However, it is controversial whether they have a totally diffusive nature or include some contributions from yet-unresolved, dim gamma-ray sources. \cite{Fang_2021} suggest that some very energetic sources such as hypernova remnants could largely contribute to the Tibet AS$\gamma$ diffuse gamma-ray events. \cite{Vecchiotti_2022} also demonstrate that the Tibet AS$\gamma$ diffuse gamma-ray events can be explained with the truly diffuse emission plus the emission from unresolved pulsar-wind nebulae too dim to be detected, the latter of which could make a significant contribution.

The possible pictures above can be tested experimentally with the LHAASO observatory which is now steadily operating and surveying gamma rays from the northern sky in the sub-PeV energy range with a sensitivity one order of magnitude better than the other existing gamma-ray observatories \citep{LHAASO100TeV, doi:10.1126/science.abg5137}. Considering the excellent sensitivity of LHAASO, one can naturally expect their detection of many gamma-ray sources that the Tibet AS$\gamma$ experiment could not detect in the directions the Tibet AS$\gamma$ diffuse gamma-ray events come from. Furthermore, the LHAASO collaboration recently reported their first catalog (1LHAASO catalog) of gamma-ray sources detected in $1\, {\rm TeV}<E<25\, {\rm TeV}$ and/or $E > 25\, {\rm TeV}$, and 43 out of the sources found in the latter energy range are also detected above $100\, {\rm TeV}$ with more than $4\, {\sigma}$ levels \citep{1LHAASOCatalog}. The detection of the sub-PeV gamma-ray sources gives us a good opportunity to consider whether these sources contribute to the Tibet AS$\gamma$ diffuse gamma-ray events.

\section{Results and discussions} \label{sec:result}

\subsection{Tibet AS$\gamma$ Galactic diffuse gamma-ray emission events}
This study focuses on the 23 gamma-ray-like diffuse events above $398\, {\rm TeV}$ detected by the Tibet AS$\gamma$ experiment in the Galactic latitudinal range of $|b| < 10^{\circ}$, as reported in \cite{TibetDiffuse}; hereafter let us simply call these events the {\it Tibet AS$\gamma$ Galactic diffuse gamma-ray emission (T-GDE) events}. These events are found in the data taken by the observatory during a live time of $719$ days from February 2014 to May 2017. The Tibet AS$\gamma$ experiment consists of a surface air shower array covering a geometrical area of $65\, 700\, {\rm m}^2$ and an underground muon detector array with a $3400\, {\rm m}^2$ area. The latter counts the number of muons in air showers, leading to powerful discrimination of primary gamma rays and cosmic rays. \cite{TibetDiffuse} found the T-GDE events after the event selection with the tight muon cut which realizes the cosmic-ray survival ratio of $10^{-6}$ and the gamma-ray survival ratio of $30\%$ above $398\, {\rm TeV}$. Note that the 23 events include $2.73$ background cosmic-ray events and four gamma-ray-like events detected within $4^{\circ}$ from the center of the Cygnus Cocoon; see \cite{TibetDiffuse}.

\subsection{Positions of the T-GDE events and the sub-PeV LHAASO sources}
\cite{Vecchiotti_2022} propose that $\simeq 50\%$ of the diffuse gamma-ray flux above $398\, {\rm TeV}$ measured by the Tibet AS$\gamma$ experiment could be the contributions from gamma-ray sources unresolved by the Tibet AS$\gamma$ experiment. This means that $\sim 50\%$ of the T-GDE events come from unresolved sources. If the 1LHAASO catalog sources detected above 100 TeV (hereafter simply called {\it sub-PeV LHAASO sources}) are the origins of the T-GDE events, about 10 sources should overlap the T-GDE events under the picture proposed by \cite{Vecchiotti_2022}. Is that the case?

Figure \ref{fig1} shows the incoming directions of the T-GDE events (red points) and the positions and extensions of the sub-PeV LHAASO sources (blue circles; \citealp{1LHAASOCatalog}). Assuming a two-dimensional Gaussian morphology, the extensions of $95\%$ containment radii are calculated from the published extensions of the $39\%$ containment radii \citep{1LHAASOCatalog}. For point-like sources, the $95\%$ upper limits on the extensions are shown. Surprisingly, no T-GDE events have their arrival directions within the extensions of the sub-PeV LHAASO sources. This means that none of the sub-PeV LHAASO sources is the origin of the T-GDE events. 

\subsection{Statistical consistency of the observations by the Tibet AS$\gamma$ experiment and LHAASO}\label{sec:potentialLHAASO}
\begin{figure}
  \centering
  \includegraphics[scale=1.0]{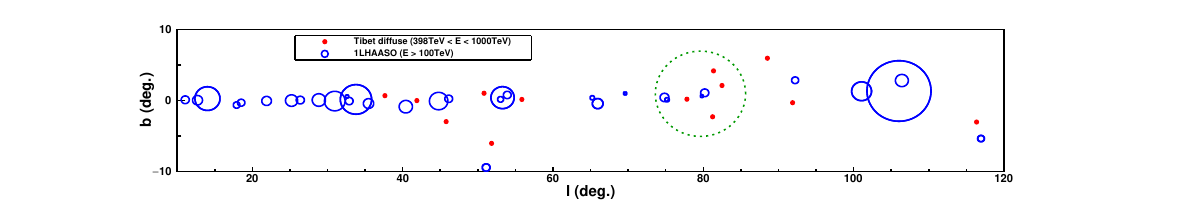}
  \includegraphics[scale=1.0]{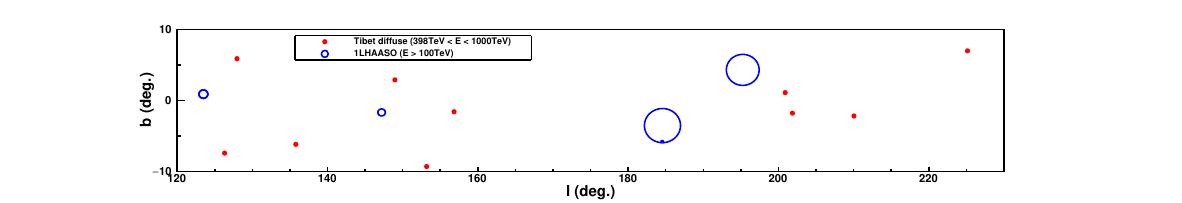}
  \caption{Incoming directions of the Galactic diffuse gamma-ray events above $398\, {\rm TeV}$ observed by the Tibet AS$\gamma$ experiment (red points) and the positions and extensions of the sub-PeV LHAASO sources: the gamma-ray sources detected by LHAASO above $100\, {\rm TeV}$ (blue circles; \cite{1LHAASOCatalog}) in the Galactic coordinates. The extensions of the sub-PeV LHAASO sources have $95\%$ containment radii assuming a two-dimensional Gaussian morphology. For point-like sources, the $95\%$ upper limits on the extensions are shown. The green dashed circle with a $6^{\circ}$ radius encloses the Cygnus region centered at $(l,\, b) = (79\fdg 62,\, 0\fdg 96)$, the position of ${\rm HAWC J}2030+409$ \citep{HAWCCygnus}. Note that some of the sub-PeV LHAASO sources have $|b| > 10^{\circ}$ and thus are not shown in the plot.}
  \label{fig1}
\end{figure}

One may think of the accidental positional overlap of the T-GDE events with the sub-PeV LHAASO sources. The sum of the extensions (in terms of $95\%$ containment radii) of the sub-PeV LHAASO sources located in the area of interest shown in Figure \ref{fig1} covers $\simeq 3.7\%$ of the area of interest, and the expected number of the T-GDE events that accidentally overlap the sub-PeV LHAASO sources is estimated to be $\simeq 0.037\times 23 = 0.86$. Therefore, it is natural to see no accidental overlap between the T-GDE events and the sub-PeV LHAASO sources, as observed.

\begin{figure}
  \centering
  \includegraphics[scale=0.4]{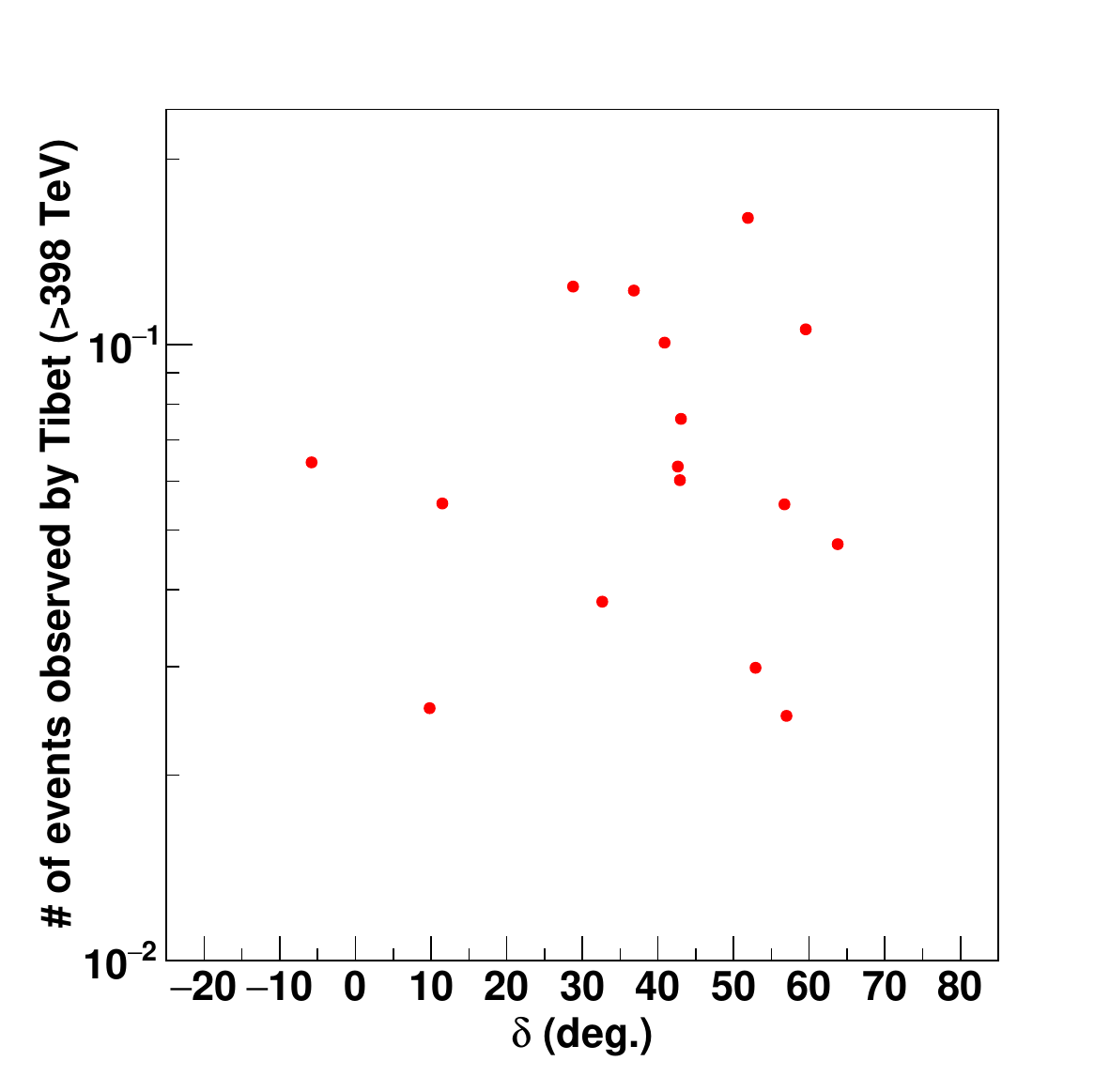}
  \caption{Expected number of gamma-ray events above 398 TeV that would be observed by the Tibet AS$\gamma$ experiment from the 22 newly-reported sub-PeV LHAASO sources. For more details, see the main text.}
  \label{fig2}
\end{figure}

In the Tibet AS$\gamma$ diffuse-gamma-ray analysis performed by \cite{TibetDiffuse}, they masked the gamma-ray sources listed in the TeVCat catalog as of 2021. Twenty out of the 43 sub-PeV LHAASO sources are associated with the TeVCat catalog sources masked in the analysis of \cite{TibetDiffuse}, while the remaining 22 sources are not; hereafter such sources are called {\it newly-reported sub-PeV LHAASO sources}, listed in Table \ref{tab:1}. The statistical consistency of the non-detection of gamma-ray events by the Tibet AS$\gamma$ experiment from these sources should be studied. The number of gamma-ray events $n$ above 398 TeV from each of the newly-reported sub-PeV LHAASO sources can be estimated as
\begin{equation} \label{eq:prob}
  n = F(>398\, {\rm TeV}) \times \Bigg( \frac{\displaystyle \int_{\theta < 40^{\circ}}\, D\, {\rm cos}\, \theta(t)\, {\rm d}t}{\displaystyle \int_{\theta < 40^{\circ}} \, {\rm d}t} \Bigg) \times S_{\rm MD} \times T_{\theta<40^{\circ}}.
\end{equation}
Here $F(>398\, {\rm TeV})$ is the integral gamma-ray flux above $398\, {\rm TeV}$ of the source. $F(>398\, {\rm TeV})$ is estimated from differential flux and spectral index determined from the LHAASO KM2A measurement above 25 TeV, because \cite{1LHAASOCatalog} do not present the integral gamma-ray fluxes above $398\, {\rm TeV}$ for the sub-PeV LHAASO sources. $D$ $(= 65\,700\, {\rm m}^2)$ is the geometrical area of the Tibet AS$\gamma$ experiment and $S_{\rm MD}$ $(=0.3)$ is the survival ratio of gamma rays above 398 TeV due to the event selection using the underground muon detector array \citep{TibetDiffuse}. The time integrals are performed for the period in a sidereal day in which a source of interest resides in the zenith-angle ($\theta$) range within $40^{\circ}$ employed in the Tibet AS$\gamma$ data analysis. The term enclosed with the parenthesis thus calculates the effective detection area for gamma-ray events by integrating the geometrical area of the AS array projected to the time-dependent source direction. $T_{\theta<40^{\circ}}$ is the total period in which the source resides in $\theta < 40^{\circ}$ over the live time (719 days) of the dataset used in the Tibet AS$\gamma$ data analysis. The results of the $n$'s for the newly-reported sub-PeV LHAASO sources are shown in Figure \ref{fig2} as a function of declination $\delta$. Since the newly-reported sub-PeV LHAASO sources with $\delta < -10^{\circ}$ or $\delta > 70^{\circ}$ do not culminate in $\theta < 40^{\circ}$ at the site of the Tibet AS$\gamma$ experiment ($30\fdg102{\rm N}$), they have $n=0$ and are not shown in the figure. The sum of the $n$'s of the newly-reported sub-PeV LHAASO sources is $1.2$, so the non-detection of gamma-ray events by the Tibet AS$\gamma$ experiment from the newly-reported sub-PeV LHAASO sources is statistically consistent. Therefore, our result currently supports the diffusive nature of the T-GDE events, although the potential origin of the T-GDE events in the gamma-ray sources yet unresolved in the sub-PeV energy range cannot be ruled out.

\begin{deluxetable*}{lrl}
\tablenum{1}
\tablecaption{List of the newly-reported sub-PeV LHAASO sources\label{tab:1}}
\tablehead{\colhead{Source Name} & \colhead{${\delta} (^{\circ})$} & \colhead{$n$}}
\startdata
1LHAASO J0007+5659u & $57.00$ & $2.5\times 10^{-2}$ \\
1LHAASO J0007+7303u & $73.07$ & $0$ \\
1LHAASO J0056+6346u & $63.77$ & $4.7\times 10^{-2}$ \\
1LHAASO J0206+4302u & $43.05$ & $7.5\times 10^{-2}$ \\
1LHAASO J0212+4254u & $42.91$ & $6.0\times 10^{-2}$ \\
1LHAASO J0216+4237u & $42.63$ & $6.3\times 10^{-2}$ \\
1LHAASO J0343+5254u & $52.91$ & $3.0\times 10^{-2}$ \\
1LHAASO J1740+0948u & $9.81$ & $2.6\times 10^{-2}$ \\
1LHAASO J1809-1918u & $-19.30$ & $0$ \\
1LHAASO J1814-1719u & $-17.89$ & $0$ \\
1LHAASO J1814-1636u & $-16.62$ & $0$ \\
1LHAASO J1825-1256u & $-12.94$ & $0$ \\
1LHAASO J1825-1337u & $-13.63$ & $0$ \\
1LHAASO J1839-0548u & $-5.8$ & $6.4\times 10^{-2}$ \\
1LHAASO J1959+2846u & $28.78$ & $0.12$ \\
1LHAASO J1959+1129u & $11.49$ & $5.5\times 10^{-2}$ \\
1LHAASO J2002+3244u & $32.64$ & $3.8\times 10^{-2}$ \\
1LHAASO J2020+3649u & $36.82$ & $0.12$ \\
1LHAASO J2031+4052u & $40.88$ & $0.10$ \\
1LHAASO J2108+5153u & $51.90$ & $0.16$ \\
1LHAASO J2200+5643u & $56.73$ & $5.5\times 10^{-2}$ \\
1LHAASO J2229+5927u & $59.55$ & $0.11$ \\
\enddata
\tablecomments{$\delta$, declination of the source; $n$, the expected number of gamma-ray events that would be observed by the Tibet AS$\gamma$ experiment from the source during the live time, calculated from Equation \eqref{eq:prob}. The sources outside the field of view of the Tibet AS$\gamma$ experiment have $n=0$; see the main text.}
\end{deluxetable*}

\subsection{Positional correlation between the T-GDE events and gamma-ray sources other than the sub-PeV LHAASO sources}
This study checks overlaps between the T-GDE events and existing gamma-ray sources other than the sub-PeV LHAASO sources to study the potential source origin of the T-GDE events. First, the positional correlation between the T-GDE events and the total 47 1LHAASO catalog sources detected only below 100 TeV is studied. The circular region with a radius of $0.5^{\circ}$ centered at each of the 47 1LHAASO catalog sources is considered and the T-GDE events within these circular regions are seached. The radius of the circular regions is the same as that employed for the masked regions in the Tibet AS$\gamma$ diffuse-gamma-ray analysis \citep{TibetDiffuse}. As a result, it is found that only one T-GDE event named TASG-D01-025 with the arrival direction of $(\alpha_{\rm J2000},\, \delta_{\rm J2000}) = (286.96^{\circ},\, 7.96^{\circ})$ (see the Supplemental Material of \cite{TibetDiffuse}) is within the circular region of 1LHAASO J1907$+$0826 which is detected only in $1\,{\rm TeV}<E<25\,{\rm TeV}$ by the LHAASO Water Cherenkov Detector Array. The fact of the one overlap is statistically consistent with an expected accidental overlap (0.19 events).

Furthermore, overlaps between the T-GDE events and the gamma-ray sources listed in the latest TeVCat catalog \citep{TeVCat} are checked. Out of the total 273 sources listed in the latest TeVCat catalog \footnote{As of 7th December 2023}, there are 92 sources in the sky region displayed in Figure \ref{fig1} after removing extragalactic sources. Note that the 92 TeVCat catalog sources include the sources masked in the Tibet AS$\gamma$ diffuse-gamma-ray analysis \citep{TibetDiffuse}, while they do not include the 1LHAASO catalog sources. The circular region with a radius of $0.5^{\circ}$ centered at each of the 92 TeVCat catalog sources is considered and the T-GDE events within these circular regions are searched. As a result, there is no overlap between the T-GDE events and the TeVCat catalog sources, and it is statistically consistent with an expected accidental overlap (0.38 events). The results presented in this section further support the diffusive nature of the T-GDE events.

\section{Conclusion} \label{sec:con}
Our study shows that the 43 1LHAASO catalog sources detected above $100\, {\rm TeV}$ (sub-PeV LHAASO sources) are not the origins of the 23 T-GDE events above $398\, {\rm TeV}$ from the fact that no T-GDE events have their arrival directions within the extensions of the sources. The number of accidental overlaps between the T-GDE events and the sub-PeV LHAASO sources is estimated to be $0.86$, and the expected number of gamma-ray events from the 22 newly-reported sub-PeV LHAASO sources is $1.2$, both ensuring the statistical consistency of no overlap between the T-GDE events and the sub-PeV LHAASO sources. No overlap between the T-GDE events and the sub-PeV LHAASO sources currently supports the diffusive nature of the T-GDE events, although the potential origin of the T-GDE events in the gamma-ray sources yet unresolved in the sub-PeV energy range cannot be ruled out.



\begin{acknowledgments}
 This work is supported in part by Grants-in-Aid for Scientific Research from the Japan Society for the Promotion of Science in Japan, the joint research program of the Institute for Cosmic Ray Research (ICRR), the University of Tokyo, and the use of the computer system of ICRR. This work is also supported by the National Natural Science Foundation of China under Grants No. 12227804, and the Key Laboratory of Particle Astrophysics, Institute of High Energy Physics, CAS.
\end{acknowledgments}

\end{document}